\documentclass{article}
\usepackage{axodraw}
\usepackage{amssymb,amsfonts,amsmath}
\def\espr{Q^{2} \left( 2 \xi - 1 \right) + 2 W^{2} \xi}
\def\newesp{2 S^{2} -2S \left( Q^{2} + W^{2} \right) + \left(
Q^{2} + W^{2} \right)^{2}}
\def\rad{\sqrt{1 - \frac{4 m_{c}^{2}}{\espr}}}
\def\cbcuno{\left| c \bar{c} \right>_{1}}
\def\cbcotto{\left| c \bar{c} \right>_{8}}
\begin{document}
\large
\title{Inelastic \( J/\psi \) electroproduction in e-p collisions}
\author{P. Balbi, A.Giovannini \\
Theoretical Physics Department, University of Torino}
\maketitle
Report-no: DFTT 39/02 \\
\ \ \ \\
Following HERA experiments our interest is focused, in the present paper,
on charmonium production in lepton-proton collisions. \\
Inelastic \( J/\psi \) electroproduction is quite difficult to be measured
due to the low rates of events; almost all experimental studies
are indeed performed in photoproduction regime. \\
\( J/\psi \) resonance is considered as the product of the hadronization of a
\( c \bar{c} \) pair generated via boson-gluon fusion. \\
The leading order contribution to the differential cross section,
with respect to virtual photon-proton center of mass energy \( W \) and
transferred four-momentum squared \( Q^{2} \), is written as the sum of two
terms related to a colour singlet and a colour octet \( c \bar{c} \) pair
production respectively. \\
This expression contains two-gluons contributions and depends on three
parameters; two of them are related to the hadronization fractions in
charmonium states for the \( c \bar{c} \) pair (colour singlet and colour
octet), while the third parameter takes into accout possible ``hot-spot''
effects. \\
The values of these parameters are estimated by mean of a fit to the
experimental data collected by H1 collaboration at HERA. \\
By using different gluon densities, different sets of parameters are obtained,
always with good fits. \\
\newpage
Hidden charm production in e-p collisions at HERA is an intriguing
subject both from the theoretical and from experimental point of view. \\ 
Available experimental data concern mainly charmonium photoproduction which
corresponds to \( Q^{2} \to 0 \) limit (\( q^{2} = -Q^{2} \) is the
transferred four-momentum squared) \cite{photo1} \cite{photo2} \cite{photo3}
\cite{photo4}. \\
In view of the low rates of events (cross sections decreasing with respect to
\( Q^{2} \)) only one sample of data (taken by H1 detector) on \( J/\psi \)
electroproduction is available \cite{elepro}. \\
It should be pointed out that \( e  \ p \rightarrow e \ J/\psi  \ X \)
photoproduction cross sections are characterized by one perturbative mass
scale, namely the charm quark mass \( m_{c} \). \\
These cross sections are found to be dominated by diffractive processes
whose hallmark is the low value for \( X \) mass \( M_{X} \). \\
Coming to electroproduction events there are two perturbative mass scales
which come into play, \( m_{c} \)  and \( Q \). \\
Either diffractive photoproduction or electroproduction experimental rates
have been recently understood in the framework of perturbative QCD
\vspace{0.5 in}
(pQCD) \cite{diff1} \cite{diff2} \cite{diff3}. \\
The main aim of this paper is the analysis of inelastic \( J /\psi \)
electroproduction; the true inelastic events are characterized by a
\( M_{X} \) value much higher than the diffractive ones. \\
The inelastic contributions to the inclusive cross sections can be therefore
isolated by mean of a cutoff \( M_{min} \) on the values for \( M_{X} \) (H1
coll. suggests \( M_{min} = 10 \frac{GeV}{c^{2}} \) \cite{elepro}). \\
H1 experimental points cover the kinematical range
\[ 2 < Q^{2} < 80 \ GeV^{2} \]
so that electroweak corrections (\( Z_{0} \) exchange) can be neglected and
one can work in single-photon-exchange approximation. \\
HERA results on open charm production in \( e^{+} p \) DIS suggest to
describe also inelastic charmonium electroproduction in pQCD by \( c \bar{c}
\) pairs formation via boson-gluon fusion (BGF) mechanism given by the
following partonic subprocess :
\[ \gamma^{*} \ g \rightarrow c \ \bar{c} \ X \]
That is to say any intrinsic charm contribution to the proton wavefunction is
neglected. \\
\begin{center} \begin{picture}(-200,-25)(200,90)
\Line(100,50)(122,61)
\Gluon(20,90)(78,61){2.5}{6}
\Line(78,61)(100,50)
\ArrowLine(122,61)(180,90)
\Photon(20,10)(100,50){2.5}{10}
\ArrowLine(190,65)(120,55)
\CCirc(100,50){30}{}{}
\Text(20,0)[uc]{\( \gamma^{*} \)}
\Text(20,100)[dc]{\( g \)}
\Text(190,95)[dc]{\( c \)}
\Text(200.5,65.15)[uc]{\( \bar{c} \)}
\Line(129.5,45)(175,30)
\Line(128,39)(165,15)
\Line(175,30)(180,37.5)
\Line(165,15)(160,7.5)
\Line(160,7.5)(184.5,15)
\Line(180,37.5)(184.5,15)
\Text(192,14)[dc]{\( X \)}
\Text(106,-25)[cc]{Figure 1: BGF mechanism}
\end{picture} \end{center}
\vspace{1.81 in}
The Born contribution is given by the following partonic diagrams:
\begin{center} \begin{picture}(350,180)(10,0)
\Gluon(20,130)(70,130){3}{7.5}
\Gluon(200,130)(250,130){3}{7.5}
\ArrowLine(160,166)(70,130)
\ArrowLine(70,130)(90,122)
\ArrowLine(20,30)(90,58)
\ArrowLine(90,58)(160,30)
\ArrowLine(250,130)(340,94)
\ArrowLine(270,138)(250,130)
\ArrowLine(200,30)(270,58)
\ArrowLine(270,58)(340,30)
\ArrowLine(340,170)(270,138)
\ArrowLine(90,122)(160,94)
\Photon(90,122)(90,58){3}{7.5}
\Photon(270,58)(270,118){3}{7.5}
\Photon(270,126)(270,138){3}{1.5}
\CArc(270,122)(4,-90,90)
\Vertex(70,130){1.6}
\Vertex(92,121.5){1.6}
\Vertex(90,58){1.6}
\Vertex(268,137.5){1.6}
\Vertex(270,58){1.6}
\Vertex(250,130){1.6}
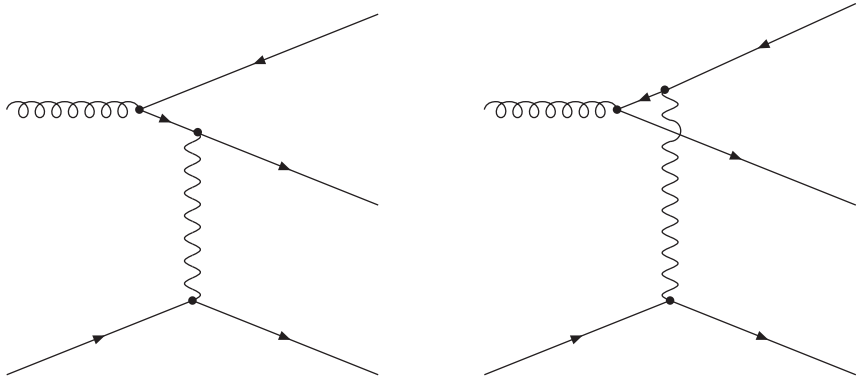
\Text(180,5)[cc]{Figure 2: Born terms for \( e \ g \rightarrow e \ c \ \bar{c}
\)}
\end{picture} \end{center}
\vspace{0.25 in}
Of course these diagrams contribute to the cross sections for production of
a \( c \bar{c} \) pair in a colour octet state (this pair is coupled to a
gluon-photon system)
\[ e \ p \rightarrow e \ \cbcotto \ X \]
The emission of an additional gluon allows the \( c \bar{c} \) pair to emerge
in a colour singlet state; so the lowest order contribution to the
\[ e \ p \rightarrow e  \ \cbcuno \ X \]
process comes from the following diagrams
\begin{center} 
\begin{picture}(360,290)(20,0)
\ArrowLine(80,250)(120,270)
\ArrowLine(120,270)(160,290)
\ArrowLine(110,235)(80,250)
\ArrowLine(160,210)(110,235)
\ArrowLine(60,160)(110,185)
\ArrowLine(110,185)(160,160)
\Gluon(30,250)(80,250){2.5}{6}
\Gluon(120,270)(160,270){-2.0}{5}
\Photon(110,185)(110,235){2.0}{6}
\Vertex(110,185){1.6}
\Vertex(110,235){1.6}
\Vertex(80,250){1.6}
\Vertex(120,270){1.6}
\ArrowLine(260,250)(340,290)
\ArrowLine(300,230)(280,240)
\ArrowLine(340,210)(300,230)
\ArrowLine(280,240)(260,250)
\ArrowLine(260,160)(300,180)
\ArrowLine(300,180)(340,160)
\Gluon(210,250)(260,250){2.5}{6}
\Gluon(280,240)(330,240){2.5}{6}
\Photon(300,180)(300,230){2.0}{6}
\Vertex(300,180){1.6}
\Vertex(300,230){1.6}
\Vertex(260,250){1.6}
\Vertex(280,240){1.6}
\ArrowLine(80,90)(160,130)
\ArrowLine(110,75)(80,90)
\ArrowLine(135,62.5)(110,75)
\ArrowLine(160,50)(135,62.5)
\ArrowLine(60,0)(110,25)
\ArrowLine(110,25)(160,0)
\Gluon(30,90)(80,90){2.5}{6}
\Gluon(135,62.5)(165,62.5){2.0}{4}
\Photon(110,25)(110,75){2.0}{6}
\Vertex(110,25){1.6}
\Vertex(110,75){1.6}
\Vertex(80,90){1.6}
\Vertex(135,62.5){1.6}
\ArrowLine(260,90)(290,105)
\ArrowLine(290,105)(315,117.5)
\ArrowLine(315,117.5)(340,130)
\ArrowLine(340,50)(260,90)
\ArrowLine(240,0)(290,25)
\ArrowLine(290,25)(340,0)
\Gluon(210,90)(260,90){2.5}{6}
\Gluon(315,117.5)(345,117.5){-2.0}{4}
\Photon(290,80)(290,105){2.0}{3.5}
\Photon(290,25)(290,70){2.0}{5.5}
\CArc(290,75)(5,-90,90)
\Vertex(290,25){1.6}
\Vertex(290,105){1.6}
\Vertex(315,117.5){1.6}
\Vertex(260,90){1.6}
\ArrowLine(80,-70)(100,-60)
\ArrowLine(100,-60)(120,-50)
\ArrowLine(160,-110)(80,-70)
\ArrowLine(120,-50)(160,-30)
\ArrowLine(80,-150)(120,-130)
\ArrowLine(120,-130)(160,-150)
\Gluon(30,-70)(80,-70){2.5}{6}
\Gluon(100,-60)(115,-60){-2.5}{2.0}
\Gluon(125,-60)(160,-60){-2.5}{4.0}
\Photon(120,-130)(120,-95){2.0}{5.5}
\Photon(120,-85)(120,-50){2.0}{3.5}
\CArc(120,-90)(5,-90,90)
\CArc(120,-60)(5,0,180)
\Vertex(120,-130){1.6}
\Vertex(120,-50){1.6}
\Vertex(80,-70){1.6}
\Vertex(100,-60){1.6}
\ArrowLine(260,-70)(290,-55)
\ArrowLine(290,-55)(340,-30)
\ArrowLine(300,-90)(260,-70)
\ArrowLine(340,-110)(300,-90)
\ArrowLine(240,-150)(290,-125)
\ArrowLine(290,-125)(340,-150)
\Gluon(210,-70)(260,-70){2.5}{6}
\Gluon(310,-95)(345,-95){2.5}{4.5}
\Photon(290,-125)(290,-90){2.0}{4.5}
\Photon(290,-80)(290,-55){2.0}{3.5}
\CArc(290,-85)(5,-90,90)
\Text(187.5,-175)[cc]{Figure 3: Lowest order contributions to \( e \ g
\rightarrow e \ g \ c \ \bar{c} \)}
\Vertex(290,-125){1.6}
\Vertex(290,-55){1.6}
\Vertex(260,-70){1.6}
\Vertex(310,-95){1.6}
\end{picture}
\end{center}
\vspace{2.59 in}
Diagrams in Figure 3 contribute both to \( e \ p \rightarrow e \ \cbcuno \ X
\) and to \( e \ p \rightarrow e \ \cbcotto \ X \). \\
It should be pointed out that the two other diagrams shown in figure 4
contribute to \( e \ p \rightarrow e \ \left| c \bar{c} \right>_{8} \ X \)
process at the same perturbative order. \\
\begin{center}   \begin{picture}(-325,-40)(310,90)
\ArrowLine(75,69)(130,96.5)
\ArrowLine(130,41.5)(95,59)
\ArrowLine(95,59)(75,69)
\ArrowLine(10,-13.5)(65,14)
\ArrowLine(65,14)(120,-13.5)
\ArrowLine(295,96.5)(240,69)
\ArrowLine(260,59)(295,41.5)
\ArrowLine(240,69)(260,59)
\ArrowLine(175,-13.5)(230,14)
\ArrowLine(230,14)(285,-13.5)
\ArrowLine(175,-13.5)(230,14)
\ArrowLine(230,14)(285,-13.5)
\Gluon(0,69)(75,69){2.0}{8.0}
\Gluon(37.5,69)(73.5,87){2.0}{4.5}
\Gluon(165,69)(240,69){2.0}{8.0}
\Gluon(202.5,69)(238.5,87){2.0}{4.5}
\Photon(65,14)(95,59){2.0}{6.0}
\Photon(230,14)(260,59){2.0}{6.0}
\Vertex(64,14){1.6}
\Vertex(95,59){1.6}
\Vertex(75,69){1.6}
\Vertex(37.5,69){1.6}
\Vertex(229,14){1.6}
\Vertex(260,59){1.6}
\Vertex(240,69){1.6}
\Vertex(202.5,69){1.6}
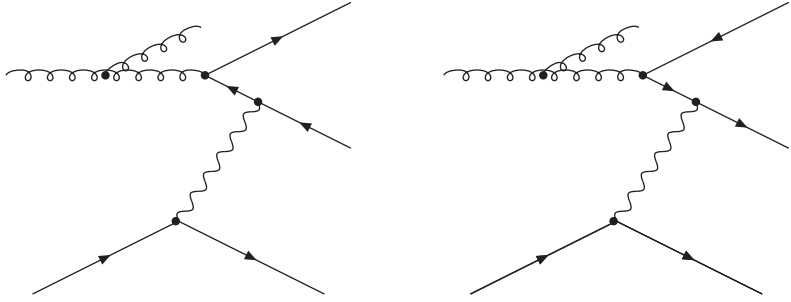
\Text(147.5,-40)[cc]{Figure 4: Lowest order contributions to \( e \ g
\rightarrow e \ g \ \left| c \bar{c} \right>_{8} \)}
\end{picture}   \end{center}
\vspace{2.1 in}
If we examine the overall lowest order contribution to the squared matrix
element for \( e \ g \rightarrow e \ g \ c \ \bar{c} \) transition, we easily
find that it enjoys the following colour decomposition:
\begin{equation}
\overline{\sum} \left| M_{LO} \left( e \ g \rightarrow e \ g \ c \ \bar{c}
\right) \right|^{2} = S_{QED} T \left( F \right) C_{2} \left( F \right) +
S_{NA} C_{2} \left( A \right)
\end{equation}
where the symbol \( \overline{\sum} \) means that we are averaging over colour
and polarization indexes for incoming particles and summing over outgoing
states. \\
\( S_{QED} \) and \( S_{NA} \) are the ``space-time'' coefficients; \( S_{QED}
\) takes contribution only from the diagrams in figure 3, while \( S_{NA} \)
includes also diagrams in figure 4. \\
The colour factors are given, as usual, by
\[ C_{2} \left( F \right) = \frac{N^{2} - 1}{2N} = \frac{4}{3} \ , \ T \left(
F \right) = \frac{1}{2} \ , \ C_{2} \left( A \right) = N = 3 \]
The lowest order terms of the gluon initiated cross sections for a colour
singlet \( c \bar{c} \) pair inclusive production \( \sigma_{g} \left( e \ p
\rightarrow e \ \cbcuno \ X \right) \) turn out to be collinear finite since
the charm quark is massive. \\
Moreover the soft divergences found in the perturbative expansions for \(
\sigma_{g} \left( e \ p \rightarrow e \ \cbcotto \ X \right) \) and \(
\sigma_{g} \left( e \ p \rightarrow e \ \cbcuno \ X \right) \) compensate
each other (order by order in perturbation theory); then the cross sections
\( \sigma_{g} \left( e \ p \rightarrow e \ c \ \bar{c} \ X \right) \) summed
over colour states for the \( c \bar{c} \) pair remain finite.
They have been explicitly computed in the literature, up to next to leading
order \cite{hf1} \cite{hf2} \cite{hf3}. \\
If we consider the colour structure of the lowest order squared matrix element
for the transition
\[ e  \ g \rightarrow e \ \cbcuno \ g \]
we find it's given by
\[ \overline{\sum} \left| M_{LO} \left( e  \ g \rightarrow e \ \cbcuno \ g
\right) \right|^{2} = S_{QED} C_{1} \ , \ C_{1} = \frac{1}{8} T \left( F
\right) C_{2} \left( F \right) \]
where \( C_{1} \) is the corresponding colour factor. \\
It's important to notice that H1 events run over the ranges
\[ 40 < W < 180 \ GeV \ , 0.2 < z \lesssim 1 \]
where \( W \) is the center of mass energy of the virtual photon-proton
system and \( z \) is the ratio of \( J/\psi \) and exhanged photon energies in
the proton rest frame (elasticity of the process). \\
It can be easily shown that the Bjorken scaling variable \( x \), in terms of 
\( W \) and \( Q^{2} \), is given by \( x = \frac{Q^{2}}{W^{2}+ Q^{2}} \). \\
Accordingly the  experimental points lie in the low \( x \) region
\[ 6 \cdot 10^{-5} < x < 5 \cdot 10^{-2} \]
It's well known that the density of gluons \( \rho_{g} \left( \xi , \mu
\right) \), with respect to the fraction of longitudinal momentum carried \(
\left( \xi \right) \) and the factorization scale \( \left( \mu \right) \),
shows a rapid enhancement as \( \xi \) decreases. \\
For \( x \) values small enough gluon recombination effects can no longer be
neglected; in other words the BGF dynamics could involve 2 gluons
(boson-gluon-gluon fusion). \\
It should be stressed that these effects contribute to the cross sections for
inclusive colour singlet \( \cbcuno \) pair production already at the
next-to-leading order; so we have explicitly calculated (under appropriate
hypotheses) the lowest order contribution of the
\[ e \ g \ g \rightarrow e \  \cbcuno \]
partonic subprocess to the double-differential cross section with
respect to \( W \) and \( Q^{2} \) variables
\[ \frac{d^{2} \sigma \left( e \ p \rightarrow e \ \cbcuno \ X 
\right)}{dW dQ^{2}} \]
The three partonic diagrams shown in figure 5 should be taken into account
\begin{center}  \begin{picture}(-280,175)(290,150)
\ArrowLine(60,250)(60,280)
\ArrowLine(60,280)(120,310)
\ArrowLine(120,220)(80,240)
\ArrowLine(80,240)(60,250)
\ArrowLine(30,175)(80,200)
\ArrowLine(80,200)(130,175)
\ArrowLine(220,250)(220,280)
\ArrowLine(280,220)(220,250)
\ArrowLine(220,280)(240,290)
\ArrowLine(240,290)(280,310)
\ArrowLine(190,175)(240,200)
\ArrowLine(240,200)(290,175)
\ArrowLine(150,90)(140,105)
\ArrowLine(140,105)(130,120)
\ArrowLine(210,60)(150,90)
\ArrowLine(130,120)(190,150)
\ArrowLine(90,15)(140,40)
\ArrowLine(140,40)(190,15)
\Gluon(0,280)(60,280){2.0}{6.5}
\Gluon(0,250)(60,250){2.0}{6.5}
\Gluon(160,250)(220,250){2.0}{6.5}
\Gluon(160,280)(220,280){2.0}{6.5}
\Gluon(80,120)(130,120){2.0}{5.5}
\Gluon(80,90)(150,90){2.0}{7.5}
\Vertex(60,250){1.6}
\Vertex(60,280){1.6}
\Vertex(80,200){1.6}
\Vertex(80,240){1.6}
\Vertex(240,200){1.6}
\Vertex(240,290){1.6}
\Vertex(220,250){1.6}
\Vertex(220,280){1.6}
\Vertex(140,40){1.6}
\Vertex(140,105){1.6}
\Vertex(130,120){1.6}
\Vertex(150,90){1.6}
\Photon(80,200)(80,240){1.6}{4.5}
\Photon(240,200)(240,235){2.0}{3.5}
\Photon(240,245)(240,290){2.0}{5.0}
\Photon(140,95)(140,105){-2.0}{1.0}
\Photon(140,40)(140,85){2.0}{5.0}
\CArc(240,240)(5,-90,90)
\CArc(140,90)(5,90,270)
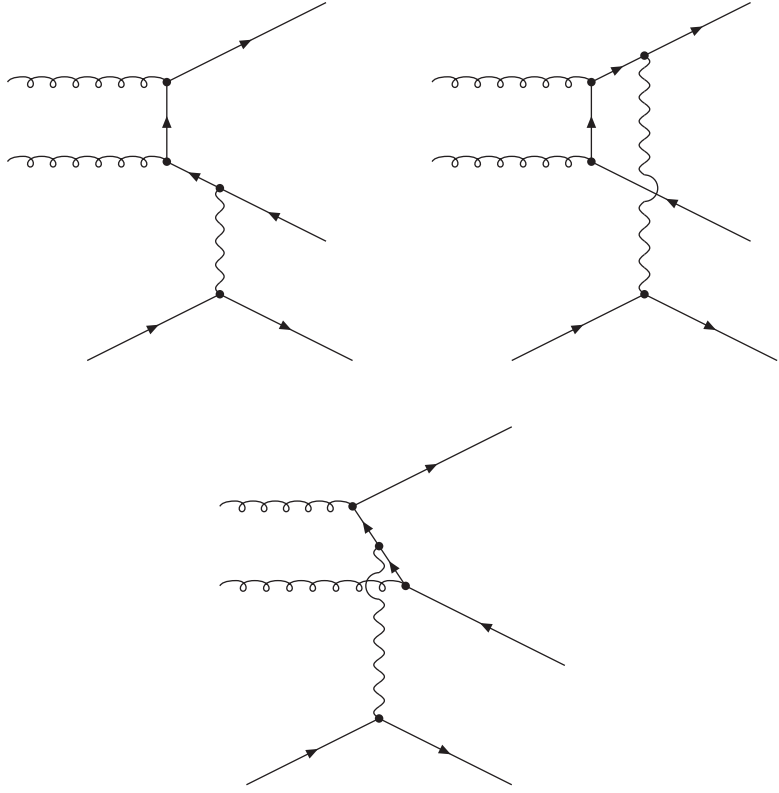
\Text(140,-10)[cc]{Figure 5: Boson-gluon-gluon fusion}
\end{picture}   \end{center}
\vspace{2.4 in}
and those obtained by reversing the orientation of the charm quark line. 6
diagrams in total. \\
Their contribution can be written as a convolution over the two-gluons density
for unit area in the proton with respect to the fractions of longitudinal
momentum carried \( \rho^{(2)}_{g} \left( \xi_{1} , \xi_{2} , \mu \right) \):
\[ \frac{d^{2} \sigma^{(LO)}_{2g} \left( e \ p \rightarrow e  \ \cbcuno \ X 
\right)}{dW dQ^{2}} = \] 
\begin{equation}
\int d \xi_{1} \int d \xi_{2} \ 
\rho^{(2)}_{g} \left( \xi_{1} , \xi_{2} , \mu \right) \hat{F}_{2g} 
\left[ \xi_{1} , \xi_{2} , Q^{2}, S, W , \alpha_{s} \left( \mu \right) \right]
\end{equation}
where \( \sqrt{S} \) is the available center of mass energy for the
electron-proton collision (at HERA \( \sqrt{S} \simeq 300 \) GeV) and
\( \alpha_{s} \left( \mu \right) \) is the strong running coupling constant. \\
In the low density limit \( \rho^{(2)}_{g} \left( \xi_{1} , \xi_{2} , \mu
\right) \) can be written as a product of single-gluon densities (the branches
of a partonic cascade evolve independently):
\[ \rho^{(2)}_{g} \left( \xi_{1} , \xi_{2} , \mu \right) = \frac{9}{8 \pi
R^{2}} \rho_{g} \left( \xi_{1} , \mu \right) \rho_{g} \left( \xi_{2} , 
\mu \right) \]
where \( R \) is the radius characteristic of the region in the proton
po- \\
pulated by gluons. \\
It's natural to identify \( R \) with the proton radius \( R \simeq 1 fm \)
(gluons are uniformely distributed inside the proton), but in the so-called
``hot spot'' effects \( R \) can be smaller (e.g \( R \) could be the radius of
a constituent quark \( R < 0.5 fm \)). \\
It should be recalled that eq. (2) is infrared divergent because of soft
singularities that occur in the limit \( \xi_{1} \to 0 \) or \( \xi_{2} \to 0
 \). \\
However gluon recombination favours gluons that are close in the rapidity
space (they have greater probability to overlap and interact). 
We make therefore the following change of variables
\[ \xi_{1} \rightarrow \xi \ , \ \xi_{2} \rightarrow e^{- \eta} \xi \]
(\( \eta \) is the difference in rapidity for the gluon pair). \\
Then eq. (2) becomes
\[\frac{d^{2} \sigma^{(LO)}_{2g} \left( e \ p \rightarrow e  \ \cbcuno \ X 
\right)}{dW dQ^{2}} = \frac{9}{8 \pi R^{2}} \int_{- \infty}^{+ \infty} \! 
d \eta e^{- \eta} \int_{\xi_{min} \left( \eta \right)}^{\xi_{max} \left( \eta
\right)} d \xi \xi \rho_{g} \left( \xi , \mu \right) \]
\begin{equation} 
\rho_{g} \left( e^{- \eta} \xi , \mu \right) {\cal{C}} \left( \left| \eta 
\right| \right) \hat{F}_{2g} \left[ \xi , \eta , Q^{2} , S , W , \alpha_{s}
\left( \mu \right) \right]
\end{equation}
\( {\cal{C}} \left( \left| \eta \right| \right) \) is here a correlation
function which implements mathematically the previous guess, while
\[ \xi_{min} \left( \eta \right) = \frac{Q^{2} + 4 m_{c}^{2}}{\left(1 + e^{- 
\eta} \right) Q^{2}} x \  , \ \xi_{max} \left( \eta \right) = \frac{1}{1 + 
e^{- \eta}} \]
Let us assume now that
\[ {\cal{C}} \left( \left| \eta \right| \right) = \delta \left( \eta \right) \]
This is, of course, a very strong hypothesis; it means that we allow
recombination  only for gluons with the same rapidity. Then one gets
\[ \frac{d^{2} \sigma^{(LO)}_{2g} \left( e \ p \rightarrow e  \ \cbcuno \ X 
\right)}{dW dQ^{2}} = \]
\begin{equation}
\frac{9}{8 \pi R^{2}} \int_{\xi_{min} \left( 0 \right)}^{
\xi_{max} \left( 0 \right)} d \xi \xi \rho_{g}^{2} \left( \xi , \mu \right)
\hat{F}_{2g} \left[ \xi , \eta=0 , Q^{2}, S , W , \alpha_{s} \left(
\mu \right) \right]
\end{equation}
Function \( \hat{F}_{2g} \left[ \xi , \eta=0 , Q^{2}, S , W , \alpha_{s}
\left( \mu \right) \right] \) in eq. (4) has been computed in pQCD,
neglecting the transverse components of incoming gluons momenta and their
interactions (free gluon wave functions). \\
Its explicit expression is rather awkward, so we decided to put it in the
appendix. \\
Eq. (4)  has been numerically integrated by using gluon densities implemented
in CERN {\bf PDFLIB} library. \\
The result turns out to be infrared safe; this is due to the hypothesis made on
\( {\cal{C}} \left( \left| \eta \right| \right) \) shape which suppresses
the initial state soft divergences. \\
Of course this contribution has to be summed to the corresponding one-gluon
term \( \frac{d^{2} \sigma^{(LO)}_{g} \left( e \ p \rightarrow e \ \cbcuno \ X
\right)}{dW dQ^{2}} \). \\
Therefore the leading order contribution to the hadronic differential cross
section for \( e \ p \rightarrow e \ J/\psi \left( z > z_{0} \right) \ X \)
process can be written as follows
\[ \frac{d^{2} \sigma^{(LO)} \left( e \ p \rightarrow e \ J/\psi \left( z >
z_{0} \right) \ X \right)}{dW dQ^{2}} = \]
\[ \frac{d^{2} \sigma^{(LO)} \left( e \ p \rightarrow e \ \cbcuno \ X
\right)}{dW dQ^{2}} \cdot {\cal{H}} \left( \cbcuno \rightarrow J/\psi \left( z
> z_{0} \right) \right) + \]
\begin{equation}
\frac{d^{2} \sigma^{(LO)} \left( e \ p \rightarrow e \ \cbcotto
\ X \right)}{dW dQ^{2}} \cdot {\cal{H}} \left( \cbcotto \rightarrow J/\psi
\left( z > z_{0} \right) \right)
\end{equation}
with
\[ \frac{d^{2} \sigma^{(LO)} \left( e \ p \rightarrow e \ \cbcuno \ X \right)}
{dW dQ^{2}} = \]
\[ \frac{d^{2} \sigma^{(LO)}_{g} \left( e \ p \rightarrow e \ \cbcuno \ X
\right)}{dW dQ^{2}} + \frac{d^{2} \sigma^{(LO)}_{2g} \left( e \ p \rightarrow
e \ \cbcuno \ X \right)}{dW dQ^{2}} \]
The term 
\[ \frac{d^{2} \sigma^{(LO)} \left( e \ p \rightarrow e \ \cbcotto \ X
\right)}{dW dQ^{2}} \] 
is the Born term which contributes to the production of
a colour octet \( c \bar{c} \) pair and \( {\cal{H}} \left( \left| c \bar{c}
\right>_{1,8} \rightarrow J/\psi \left( z > z_{0} \right) \right) \) stand for
the fraction of partonic events of type
\[ e \ p \rightarrow e \ \left| c \bar{c} \right>_{1,8} \ X \]
which hadronize in a \( J/\psi \) vector meson with elasticity \( z > z_{0} \).
\\
As we have already mentioned \( \frac{d^{2} \sigma^{(LO)}_{g} \left( e \ p
\rightarrow e \ \cbcuno \ X \right)}{dW dQ^{2}} \) is an infrared divergent
quantity (soft singularities); on this subject, the next-to-leading order
interference terms of the Born contribution with the virtual corrections
for the \( e \ g \rightarrow e \ c \ \bar{c} \) transition (which contribute
to a colour octet \( c \bar{c} \) pair production) should be considered. \\
In figure 6 are shown some of them as cut diagrams. \\
If we examine the colour structure of the interference terms we find that their
contribution to the squared matrix element can be written as:
\[ \overline{\sum} I \left( e \ g \rightarrow e \ c \ \bar{c} \right) = I_{QED}
T \left( F \right) C_{2} \left( F \right) + I_{NA} C_{2} \left( A \right) \]
where \( I_{QED} \), as before, takes contribution from ``electrodynamical''
diagrams (only quark-quark-gluon vertices), while \( I_{NA} \) contains the
contributions of ``non abelian'' diagrams. \\
Notice that \( \overline{\sum} I \left( e \ g \! \rightarrow \!
e \ c \ \bar{c} \right) \) has the same colour decomposition as \(
\overline{\sum} \left| M_{LO} \left( e \ g \rightarrow e \ g \ c \ \bar{c}
\right) \right|^{2} \); both contributions to \( \frac{d^{2} \sigma^{(NLO)}_{g}
\left( e \ p \rightarrow e \ c \ \bar{c} \ X \right)}{dW dQ^{2}} \) exhibit
soft divergences; nevertheless the last quantity remains ``infrared safe'' due
to a compensation mechanism. \\
This fact suggests to redefine \( \frac{d^{2} \sigma^{(LO)}_{g} \left( e \
p \rightarrow e \ \cbcuno \ X \right)}{dW dQ^{2}} \) by including part of the
interference terms, which enter in \( \frac{d^{2} \sigma^{(NLO)}_{g}
\left( e \ p \rightarrow e \ \cbcotto \ X \right)}{dW dQ^{2}} \), id est
\( \frac{d^{2} \tilde{\sigma}^{(LO)}_{g,1} \left( e \ p \rightarrow e \ c \
\bar{c} \ X \right)}{dW dQ^{2}} \) is defined by substituting
\[ \overline{\sum} \left| M_{LO} \left( e  \ g \rightarrow e \ \cbcuno \ g
\right) \right|^{2} = S_{QED} C_{1} \]
with
\[ \overline{\sum} \left| \tilde{M}^{(LO)}_{1} \left( e \ g \ \rightarrow e \ g
\ c \ \bar{c} \right) \right|^{2} \equiv \left( S_{QED} + I_{QED} \right)
C_{1} \]
in the expression for \( \frac{d^{2} \sigma^{(LO)}_{g} \left( e \ p
\rightarrow e \ \cbcuno \ X \right)}{dW dQ^{2}} \). \\
This operation has to be readsorbed in a redefinition of the hadroni- 
zation fractions \( {\cal{H}} \left( \left| c \bar{c} \right>_{1,8} \rightarrow
J/\psi \left( z > z_{0} \right) \right) \). \\
\begin{center}   \begin{picture}(0,-15)(400,400)
\Line(235,300)(385,300)
\Line(415,300)(565,300)
\Line(235,160)(385,160)
\Line(415,160)(565,160)
\Line(235,20)(385,20)
\Line(415,20)(565,20)
\Line(468.78,58.78)(468.78,101.22)
\DashLine(310,290)(310,400){5}
\DashLine(490,290)(490,400){5}
\DashLine(310,150)(310,260){5}
\DashLine(490,150)(490,260){5}
\DashLine(310,10)(310,120){5}
\DashLine(490,10)(490,120){5}
\CArc(262.5,365)(7.5,0,180)
\CArc(262.5,365)(7.5,180,360)
\CArc(310,365)(25,0,180)
\CArc(310,365)(25,180,360)
\CArc(490,365)(25,0,180)
\CArc(490,365)(25,180,360)
\CArc(310,220)(30,0,180)
\CArc(310,220)(30,180,360)
\CArc(490,220)(30,0,180)
\CArc(490,220)(30,180,360)
\CArc(310,80)(30,0,180)
\CArc(310,80)(30,180,360)
\CArc(490,80)(30,-135,135)
\Photon(327.18,300)(327.18,347.25){2.0}{5.0}
\Photon(292.82,300)(292.82,347.25){-2.0}{5.0}
\Photon(507.18,300)(507.18,347.25){2.0}{5.0}
\Photon(472.82,300)(472.82,347.25){-2.0}{5.0}
\Photon(288.78,160)(288.78,198.78){-2.0}{4.0}
\Photon(331.22,160)(331.22,198.78){2.0}{4.0}
\Photon(468.78,160)(468.78,198.78){-2.0}{4.0}
\Photon(511.22,160)(511.22,198.78){2.0}{4.0}
\Photon(288.78,20)(288.78,58.78){-2.0}{4.0}
\Photon(331.22,20)(331.22,58.78){2.0}{4.0}
\Photon(475,20)(475,54.05){-2.0}{4.0}
\Photon(505,20)(505,54.05){2.0}{4.0}
\Gluon(235,365)(255,365){2.0}{2.5}
\Gluon(270,365)(285,365){2.0}{2.0}
\Gluon(335,365)(385,365){2.0}{5.5}
\Gluon(415,365)(435,365){2.0}{2.5}
\Gluon(450,365)(465,365){2.0}{2.0}
\Gluon(515,365)(565,365){2.0}{5.5}
\Gluon(235,220)(280,220){2.0}{5.0}
\Gluon(340,220)(385,220){2.0}{5.0}
\Gluon(415,220)(460,220){2.0}{5.0}
\Gluon(520,220)(565,220){2.0}{5.0}
\Gluon(235,80)(280,80){2.0}{5.0}
\Gluon(340,80)(385,80){2.0}{5.0}
\Gluon(235,80)(280,80){2.0}{5.0}
\Gluon(340,80)(385,80){2.0}{5.0}
\Gluon(415,80)(460,80){2.0}{5.0}
\Gluon(520,80)(565,80){2.0}{5.0}
\GlueArc(442.5,365)(7.5,0,360){2.0}{5.0}
\GlueArc(288.78,241.22)(7.5,-127,37){2.0}{3.5}
\GlueArc(468.78,198.78)(7.5,-37,127){2.0}{3.5}
\GlueArc(280,80)(7.5,-82,82){2.0}{3.5}
\GlueArc(490,80)(30,135,225){2.0}{3.5}
\Text(400,-10)[cc]{Figure 6: \( e \ g \rightarrow e \ c \ \bar{c} \) \
interference terms}
\end{picture}   
\end{center}
\vspace{6.0 in}
\( \frac{d^{2} \tilde{\sigma}^{(LO)}_{g,1} \left( e \ p \rightarrow e \ c \
\bar{c} \ X \right)}{dW dQ^{2}} \) is found to be infrared safe in view of a
Bloch-Nordsieck compensation analogously to what happens in quantum
electrodynamics; furthermore it can be obtained from the next-to-leading
order contribution to \( \frac{d^{2} \sigma_{g} \left( e \ p \rightarrow e
 \ c \ \bar{c} \ X \right)}{dW dQ^{2}} \) (computed by Smith et al. \cite{hf1}
\cite{hf2} \cite{hf3}) simply by fixing \( C_{2} \left( A \right) = 0 \) and
multiplying by the colour factor \( \frac{1}{8} \). \\
Then we add the two-gluons contribution and we find that
\[ \frac{d^{2} \tilde{\sigma}^{(LO)}_{1} \left( e \ p \rightarrow e \ c \
\bar{c} \ X \right)}{dW dQ^{2}} = \]
\[ \frac{d^{2} \tilde{\sigma}^{(LO)}_{g,1} \left( e \ p \rightarrow e \ c \
\bar{c} \ X \right)}{dW dQ^{2}} + \frac{d^{2} \sigma^{(LO)}_{2g} \left( e \ p
\rightarrow e \ \cbcuno \ X \right)}{dW dQ^{2}} \]
Now eq. (5) can be written as follows
\[ \frac{d^{2} \sigma^{(LO)} \left( e \ p \rightarrow e \ J/\psi \left( z >
z_{0} \right) \ X \right)}{dW dQ^{2}} = \] 
\[ \frac{d^{2} \tilde{\sigma}^{(LO)}_{1} \left( e \ p \rightarrow e \ c \
\bar{c} \ X \right)}{dW dQ^{2}} \cdot \tilde{{\cal{H}}}_{1} \left( c \bar{c}
\rightarrow J/\psi \left( z > z_{0} \right) \right) + \]
\begin{equation}
\frac{d^{2} \tilde{\sigma}^{(LO)}_{8} \left( e \ p \rightarrow e \ c \ \bar{c}
\ X \right)}{dW dQ^{2}} \cdot \tilde{{\cal{H}}}_{8} \left( c \bar{c}
\rightarrow J/\psi \left( z > z_{0} \right) \right) \ ; 
\end{equation}
where
\[ \frac{d^{2} \tilde{\sigma}^{(LO)}_{8} \left( e \ p \rightarrow e \ c \
\bar{c} \ X \right)}{dW dQ^{2}} = \frac{d^{2} \sigma^{(LO)} \left( e \ p
\rightarrow e \ \cbcotto \ X \right)}{dW dQ^{2}} \]
We have already pointed out that the soft divergences in the perturbative
expansion for\( \frac{d^{2} \sigma \left( e \ p \rightarrow e \ \cbcuno \ X
\right)}{dW dQ^{2}} \) compensate, order by order, with analogous singularities
in \( \frac{d^{2} \sigma \left( e \ p \rightarrow e \ \cbcotto \ X \right)}{dW
dQ^{2}} \). \\
So the procedure introduced above, consisting in a subtraction of terms from \(
\frac{d^{2} \sigma \left( e \ p \rightarrow e \ \cbcotto \ X \right)}{dW dQ^{2}
} \) to \( \frac{d^{2} \sigma \left( e \ p \rightarrow e \ \cbcuno \ X
\right)}{dW dQ^{2}} \), in order to compensate their soft divergences, can be
extended, in principle, to all orders. \\
In this way are defined the higher order (``infrared safe'') corrections to
\( \frac{d^{2} \tilde{\sigma}_{1} \left( e \ p \rightarrow e \ c \ \bar{c} \ X
\right)}{dW dQ^{2}} \) and \( \frac{d^{2} \tilde{\sigma}_{8} \left( e \ p
\rightarrow e \ c \ \bar{c} \ X \right)}{dW dQ^{2}} \). \\
Both these quantities depend, in general, on the renormalization scale \( \mu
\) (which we set equal to the factorization scale), while the physical cross
section \( \frac{d^{2} \sigma \left( e \ p \rightarrow e \ J/\psi \left( z >
z_{0} \right) \ X \right)}{dW dQ^{2}} \) does not depend on it. \\
As a consequence \( \tilde{{\cal{H}}}_{1} \) and \( \tilde{{\cal{H}}}_{8} \)
have to depend in turns on \( \mu \) in addition to \( z_{0} \), \( W \),
\( Q^{2} \) and \( m_{c} \). \\
In the following only \( z_{0} \) dependence will be considered and
all other effects will be neglected. \\
H1 collaboration has experimentally estimated the values of the differential
cross sections \( \frac{d \sigma \left( e \ p \rightarrow e \ J/\psi \left( z >
0.2 \right) \ X \right)}{dW} \) and \( \frac{d \sigma \left( e \ p
\rightarrow e \ J/\psi \left( z > 0.2 \right) \ X \right)}{dQ^{2}} \) for
inelastic \( J/\psi \) electroproduction in \( e^{+} \ p \) collisions. \\
Therefore we have numerically integrated our leading order expression in
eq. (6), with respect to \( Q^{2} \) or \( W \). \\
The factorization/renormalization scale is set as follows
\[ \mu = \sqrt{Q^{2} + 4 m_{c}^{2}} \]
The right hand side of eq. (6) is then fitted to experimental data by using
the minimization routines in {\bf MINUIT} package of F.James and living
\( \tilde{{\cal{H}}}_{1} \), \( \tilde{{\cal{H}}}_{8} \) and \( R \) as free
parameters. \\
Different choices for the density of gluons are made, in order to compare the
fit's results. \\
The following NLO gluon densities have been used: {\bf MRS (G) (02.1995)},
{\bf CTEQ 2pM}, {\bf CTEQ 4M} and {\bf GRV 94 HO}. \\  
The results of these calculations are reported in table 1. \\
\newpage
\ \ \ \\
\ \ \ \\
\ \ \ \\
\ \ \ \\
\begin{center}
\begin{tabular}{|c|}   \hline
\multicolumn{1}{|c|}{} \\
\multicolumn{1}{|c|}{\Large \bf MRS (G) (02.1995) gluon density} \\
\multicolumn{1}{|c|}{} \\
\multicolumn{1}{|c|}{ \( R = 1.10 fm \ , \ \tilde{{\cal {H}}}_{1} \left( z_{0}
\! = \! 0.2 \right) = 0.269 \ , \tilde{\cal {{H}}}_{8} \left( z_{0} \! = \!
0.2 \right) = 2.75 \! \cdot \! 10^{-3} \)} \\
\multicolumn{1}{|c|}{ \( \chi^{2} = 6.74 \ , \ d.o.f = 10 \Rightarrow 
\chi^{2}_{red} = 0.674 \)} \\ 
\multicolumn{1}{|c|}{} \\ \hline
\multicolumn{1}{|c|}{} \\  
\multicolumn{1}{|c|}{\Large \bf CTEQ 2pM gluon density} \\
\multicolumn{1}{|c|}{} \\
\multicolumn{1}{|c|}{ \( R = 0.442 fm \ , \ \tilde{{ \cal {H}}}_{1} 
\left( z_{0} \! = \! 0.2 \right) = 0.175 \ , \tilde{{\cal {H}}}_{8} \left(
z_{0} \! = \! 0.2 \right) = 4.21 \! \cdot \! 10^{-3} \)} \\
\multicolumn{1}{|c|}{ \( \chi^{2} = 6.32 \ , \ d.o.f = 10 \Rightarrow 
\chi^{2}_{red} = 0.632 \)} \\  
\multicolumn{1}{|c|}{} \\ \hline
\multicolumn{1}{|c|}{} \\  
\multicolumn{1}{|c|}{\Large \bf CTEQ 4M gluon density} \\
\multicolumn{1}{|c|}{} \\
\multicolumn{1}{|c|}{ \( R = 0.393 fm \ , \ \tilde{{\cal {H}}}_{1} 
\left( z_{0} \! = \! 0.2 \right) = 0.109 \ , \tilde{{ \cal {H}}}_{8} \left( 
z_{0} \! = \! 0.2 \right) =  4.79 \! \cdot \! 10^{-3} \)} \\
\multicolumn{1}{|c|}{ \( \chi^{2} = 6.23 \ , \ d.o.f = 10 \Rightarrow 
\chi^{2}_{red} = 0.623 \)} \\ 
\multicolumn{1}{|c|}{} \\
\multicolumn{1}{|c|}{} \\ \hline
\multicolumn{1}{|c|}{} \\  
\multicolumn{1}{|c|}{\Large \bf GRV HO gluon density} \\
\multicolumn{1}{|c|}{} \\
\multicolumn{1}{|c|}{ \( R = 0.611 fm \ , \ \tilde{{\cal {H}}}_{1}
\left( z_{0} \! = \! 0.2 \right) = 0.160 \ , \tilde{{ \cal {H}}}_{8} \left(
z_{0} \! = \! 0.2 \right) = 5.32 \! \cdot \! 10^{-3} \)} \\
\multicolumn{1}{|c|}{ \( \chi^{2} = 6.51 \ , \ d.o.f = 10 \Rightarrow 
\chi^{2}_{red} = 0.651 \)} \\   
\multicolumn{1}{|c|}{} \\ \hline
\multicolumn{1}{c}{} \\
\multicolumn{1}{c}{Table 1: Comparison of fit results for different gluon
densities} \\
\end{tabular}
\end{center}

\vspace{0.15 in}
\newpage
In the following graphs H1 experimental points are compared with our results.
\\
\ \ \ \\
\centerline{\Large \bf MRS (G) (02.1995) gluon density}
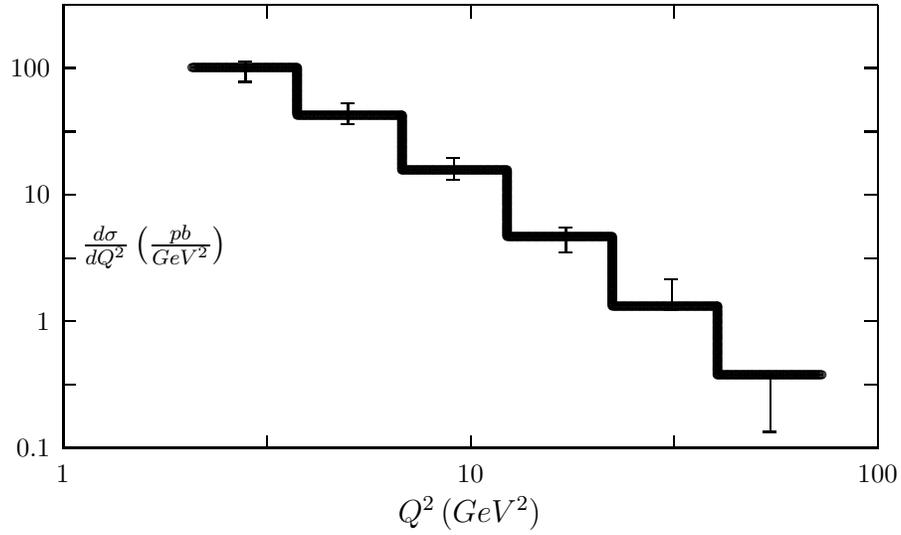
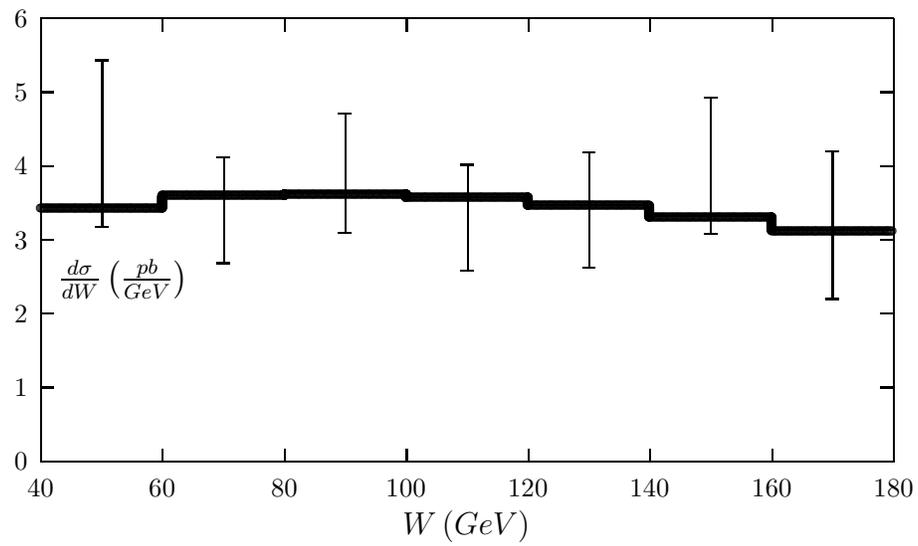
\begin{center}
\input{diag1.tex}
\ \ \ \ \ \ \ \\
\ \ \ \ \ \ \ \\
\input{diag2.tex}
\end{center}
\newpage
\ \ \ \ \ \ \ \\
\ \ \ \ \ \ \ \\
\centerline{\Large \bf CTEQ 2pM gluon density}
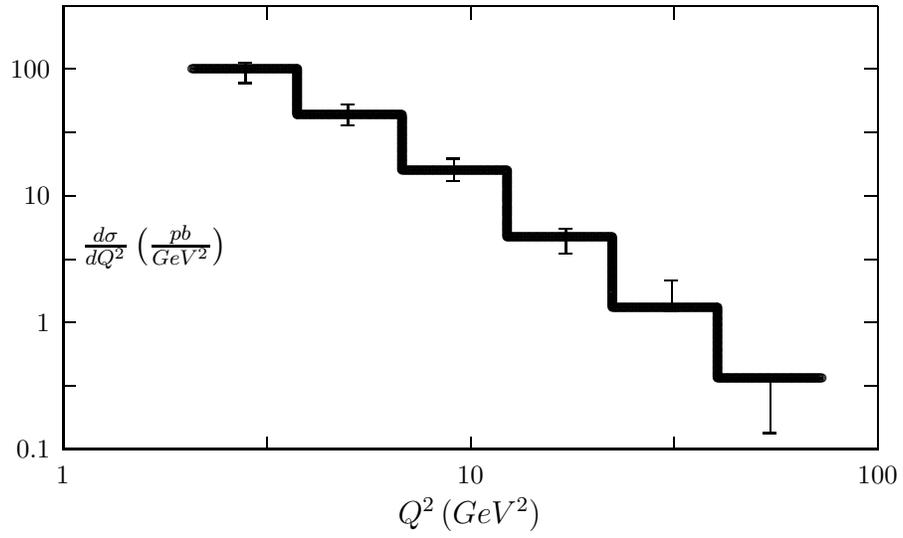
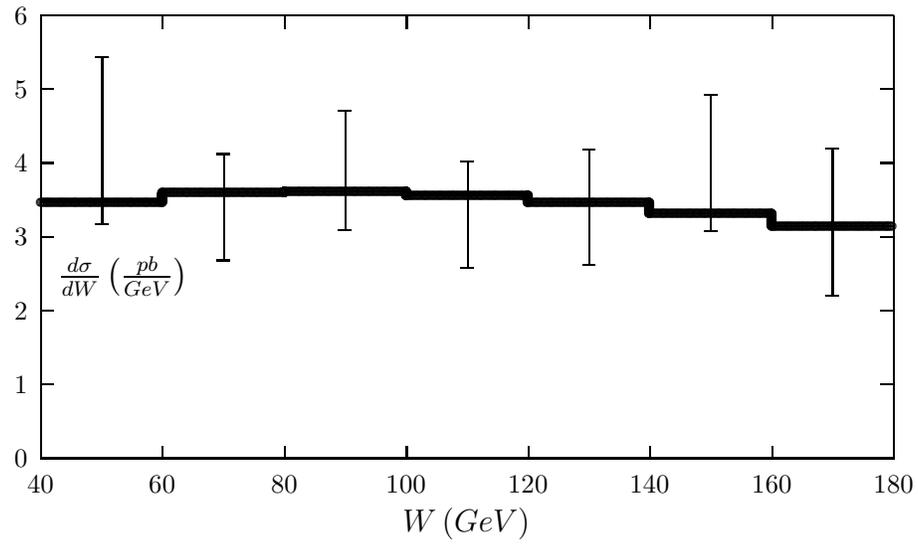
\begin{center}
\input{diag3.tex}
\ \ \ \ \ \ \\
\ \ \ \ \ \ \\
\input{diag4.tex}
\end{center}
\newpage
\ \ \ \ \ \ \ \\
\ \ \ \ \ \ \ \\
\centerline{\Large \bf CTEQ 4M gluon density}
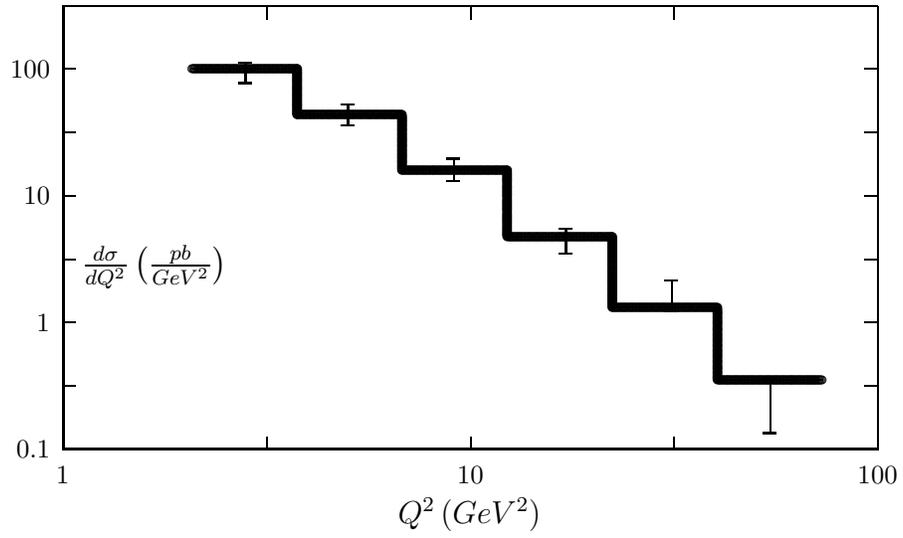
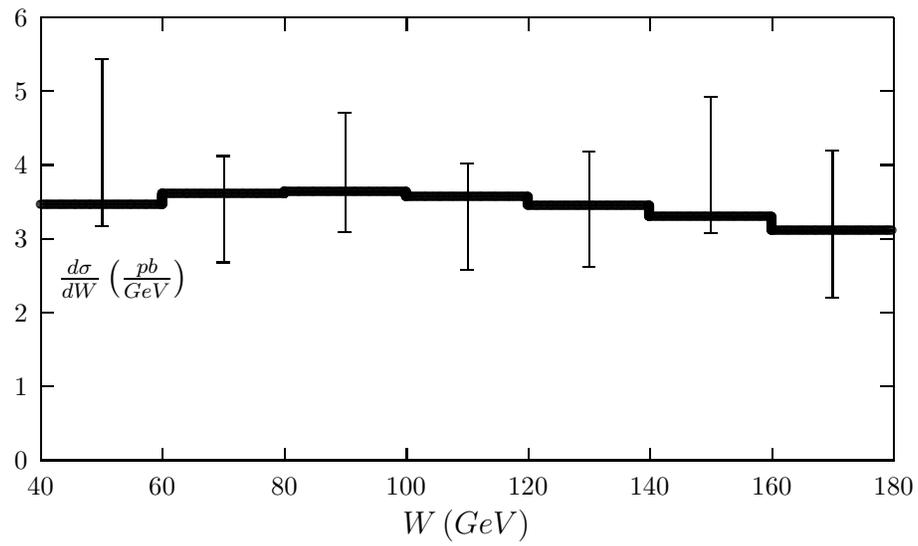
\begin{center}
\input{diag5.tex}
\ \ \ \ \ \ \\
\ \ \ \ \ \ \\
\input{diag6.tex}
\end{center}
\newpage
\ \ \ \ \ \ \ \\
\ \ \ \ \ \ \ \\
\centerline{\Large \bf GRV HO gluon density}
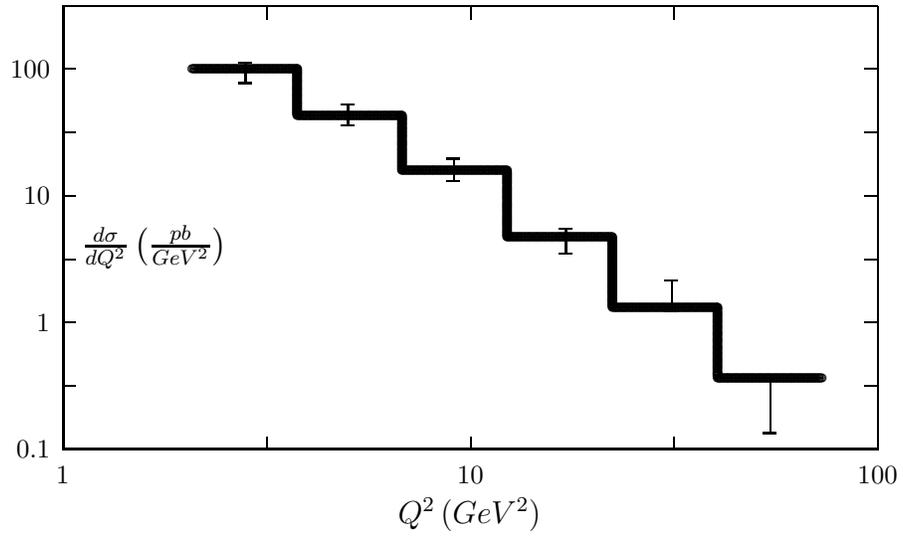
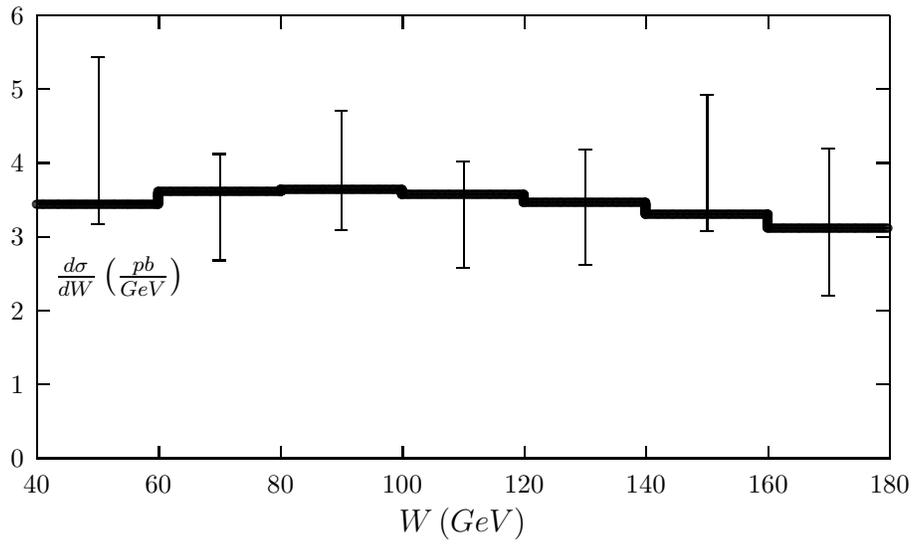
\begin{center}
\input{diag7.tex}
\ \ \ \ \ \ \\
\ \ \ \ \ \ \\
\input{diag8.tex}
\end{center}
\newpage
The errorbars are built-up summing in quadrature statistical and systematical
errors (obviously \( \chi^{2} \) has been computed taking into account only
statistical ones). \\ 
We remark that, even if these fits give values for \( \tilde{{\cal{H}}}_{8} \)
1-2 orders of magnitude smaller than for \( \tilde{{\cal{H}}}_{1} \), the
perturbative terms connected to \( \tilde{{\cal{H}}}_{1} \) are one order of
magnitude smaller than those linked with \( \tilde{{\cal{H}}}_{8} \); the
latter contribution can be up to one order of magnitude smaller than that
associated with \( \tilde{{\cal{H}}}_{1} \). \\  
The large discrepancies in the fitted parameters obtained using different
gluon densities indicate that inelastic charmonium production provides a
particularly sensitive tool to probe the gluonic structure of the hadrons;
this is due to the fact that the two-gluons contribution, which depends on the
square of the gluon density and then varies strongly with the choice of the
density itself, is of the same order of magnitude as the single-gluon
contribution (of the same perturbative order) in our kinematical range. \\
This fact suggests finally that important informations on the distribution of
gluons in the transverse space in the proton could be obtained from  hidden
charm electroproduction. \\ 
\newpage
\begin{center}
{\bf \Large Appendix}
\end{center}
\[ \hat{F}_{2g} \left[ \xi , \eta=0, Q^{2}, S , W , \alpha_{s} \left(
\mu \right) \right] = 
\frac{\pi W}{2^{4} 3^{2} \xi^{5} S^{2} Q^{2} \left( Q^{2} + W^{2} \right)^{6}}
\left\{ \alpha^{2} Q_{c}^{2} \right. \]
\[ \alpha_{s}^{2} \left( \mu \right) \left\{ \frac{1}{m_{c}^{2}} \left\{ \left[
\espr \right] \rad \right. \right. \]
\[ \left\{ 1920 m_{c}^{6} \left[ \newesp \right] + \right. \]
\[ 16 m_{c}^{4} \left\{ 4 Q^{2} \left[ 50 S^{2} -50 S \left( Q^{2} + W^{2} 
\right) + 7 \left( Q^{2} + W^{2} \right)^{2} \right] - \right. \]
\[ \left. 86 \left( Q^{2} + W^{2} \right) \left[ \newesp \right] \xi \right\} 
+ \]
\[ \left[ \newesp \right] \left[ \espr \right] \]
\[ \left[ 32 Q^{4} -40 Q^{2} \left( Q^{2} + W^{2} \right) \xi + 60
\left( Q^{2} + W^{2} \right)^{2} \xi^{2} \right] + 2 m_{c}^{2} \]
\[ \left\{ -32 Q^{4} \left[ -16 S^{2} + 16 S \left( Q^{2} + W^{2} \right) + 
\left( Q^{2} + W^{2} \right)^{2} \right] + 960 S Q^{2} \right. \]
\[ \left( Q^{2} + W^{2} \right) \left( Q^{2} + W^{2} - S \right) \xi + 44
\left( Q^{2} +  W^{2} \right)^{2} \xi^{2} \]
\[ \left. \left. \left. \left[ \newesp \right] \right\} 
\right\}  \right\} + 12 \left\{320 m_{c}^{6} \right. \]
\[ \left[ \newesp \right] + 8 m_{c}^{4} \left\{ 16 Q^{2} \right. \]
\[ \left[5 S^{2} -5S \left( Q^{2} + W^{2} \right) + \left(
Q^{2} + W^{2} \right)^{2} \right] -42 \left( Q^{2} + W^{2} \right) \]
\[ \left. \left[ \newesp \right] \xi \right\} +2 S Q^{2} \left( S- Q^{2} -
W^{2} \right) \]
\[ \left[ 32 Q^{4} -112 Q^{2} \left( Q^{2} + W^{2} \right) \xi + 100
\left( Q^{2} + W^{2} \right)^{2} \xi^{2} \right] + m_{c}^{2} \]
\[ \left\{ 256 S Q^{4} \left( S - Q^{2} - W^{2} \right) -56 Q^{2} \left(
Q^{2} + W^{2} \right) \left[10 S^{2} - \right. \right. \]
\[ \left. 10 S \left( Q^{2} + W^{2} \right) + 
\left( Q^{2} + W^{2} \right)^{2} \right] \xi + 84 \left( Q^{2} + W^{2} 
\right)^{2} \left[ \left( Q^{2} + W^{2} \right)^{2} + \right. \]
\begin{equation}
\left. \left. \left. \left. \left.\! \! \! 2 S^{2} - 2 S \left( Q^{2} + 
W^{2} \right) \right] \xi^{2} \right\} \right\} \left\{ \! \log \left\{ 
\frac{\left[ 1 + \rad \right]}{\left[ 1 - \rad \right]} \! \right\} \!
\right\} \! \right\} \! \right\}
\end{equation}
where \( \alpha \simeq \frac{1}{137} \) is the e.m fine structure constant
and \( Q_{c} = \frac{2}{3} \) represents the quark charm electric charge
(in \( e \) units). \\


\clearpage
\begin{thebibliography}{99}
\bibitem{photo1} H1 Collab., S.Aid et al.,  \emph{``Elastic and Inelastic
Photoproduction of \( J/\psi \) Mesons at HERA''}
                                            , Nucl.Phys. B472(1996) 3, 03/96
\bibitem{photo2} Zeus Collab., J.Breitweg et al.,  \emph{``Measurement of
Elastic \( J/\psi \) Photoproduction at HERA''}
                               , Z.Phys. C75(1997) 215

\bibitem{photo3} Zeus Collab., J.Breitweg et al.,  \emph{``Measurement of
inelastic \( J/\psi \) Photoproduction at HERA''}
                                     , Z.Phys C76(1997) 599

\bibitem{photo4} H1 Collab., C.Adloff et al.,  \emph{``Inelastic
Photoproduction of \( J/\psi\) Mesons at HERA''}
                                     , Eur.Phys.J. C25 (2002) 1,25-39, 05/2002


\bibitem{elepro} H1 Collab., C.Adloff et al.,  \emph{``Charmonium Production in
Deep Inelastic Scattering at HERA''}
                                        , Eur.Phys.J. C10 (1999) 373-393, 03/99


\bibitem{diff1} S.J.Brodsky et al.,   \emph{``Diffractive leptoproduction of
vector mesons in QCD''}
                                      , Phys.Rev. D50(1994) 3134

\bibitem{diff2} L.Frankfurt, W.Koepf, M.Strikman,  \emph{``Hard diffractive
electroproduction of vector mesons in QCD''}
                                                    , Phys.Rev. D54(1996) 3194

\bibitem{diff3} L.Frankfurt, W.Koepf, M.Strikman,  \emph{``Diffractive heavy
quarkonium photoproduction and electroproduction in QCD''}
                                                   , Phys.Rev. D57(1998) 512 

\bibitem{hf1} E.Laenen, S.Riemersma, J.Smith, W.L.van Neerven,  \emph{``\( O
\left( \alpha_{s} \right) \) corrections to heavy flavour inclusive
distributions in electroproduction''}
                                           , Nucl.Phys.B392: 229-250, 1993

\bibitem{hf2} E.Laenen, S.Riemersma, J.Smith, W.L.van Neerven, \emph{``Complete
\( o \left( \alpha_{s} \right) \) corrections to heavy flavour structure
functions in electroproduction''}
                                           , Nucl.Phys.B392: 162-228, 1993


\bibitem{hf3} B.W.Harris, J.Smith,  \emph{``Heavy quark correlations in deep
inelastic electroproduction''}
                                   , Nucl.Phys.B452: 109-160, 1995



\end{thebibliography}
\end{document}